\documentstyle[psfig]{mn2e}

\title[Is the soft excess in AGN real?]{Is the soft excess in AGN real?}

\author[M.~Gierli{\'n}ski and C.~Done] {Marek Gierli\'{n}ski$^{1,2}$ and Chris Done$^1$\\
$^1$Department of Physics, University of Durham, South Road, Durham DH1 3LE,UK\\
$^2$Obserwatorium Astronomiczne Uniwersytetu Jagiello\'{n}skiego,
30-244 Krak\'{o}w, Orla 171, Poland}

\date{Submitted to MNRAS}
\pagerange{\pageref{firstpage}--\pageref{lastpage}} \pubyear{2003}

\begin{document}

\def\aap{A\&A}
\def\apj{ApJ}
\def\apjl{ApJ}
\def\mnras{MNRAS}
\def\lhls{$\ell_h/\ell_s$}

\maketitle

\label{firstpage}

\begin{abstract}

We systematically analyse all publicly available {\it XMM-Newton}
spectra of radio-quiet PG quasars. The soft X-ray excess in these
objects is well modelled by an additional, cool, Compton
scattering region. However, the remarkably constant temperature
derived for this component over the whole sample requires a
puzzling fine tuning of the parameters. Instead, we propose that
the soft excess is an artifact of strong, relativistically
smeared, partially ionized absorption. The strong jump in opacity
at $\sim$0.7~keV from {\sc o\,vii}, {\sc o\,viii} and iron can
lead to an apparent soft excess below this energy, which is
trivially constant since it depends on atomic processes. This can
have a dramatic effect on the derived spectrum, which has
implications for fitting the relativistic smearing of the
reflected iron line emission from the disc.

\end{abstract}


\section{Introduction}

The broad-band optical/UV/X-ray spectra of active galactic nuclei
(AGN) consist of at least two distinct components. The big blue
bump, observed in the optical/UV band, is commonly associated with
emission from the optically thick accretion disc, while the
approximately power-law X-ray tail seen to much higher energies
requires that there is some energy dissipated outside the disc, in
an optically thin plasma (e.g. Krolik 1999). Standard models of
the accretion disc emission (Shakura \& Sunyaev 1973) predict that
the maximum temperature of the optically thick material is
$\sim$20~eV for a $10^8$~M$_\odot$ black hole accreting at the
Eddington luminosity. This disc spectrum peaks in the EUV range,
so the X-ray spectrum of AGN should be dominated instead by the
power-law coronal emission.

This contrasts with the observed soft X-ray spectra of type I AGN
(Seyfert 1's and radio quiet quasars). The intrinsic 0.2--10~keV
emission often shows a strong, broad soft excess below
$\sim$2~keV. The origin of this component is not clear. Plainly,
it is too hot to be standard disc emission from a supermassive
black hole.  One popular possibility is that there are two
Comptonizing regions. The hot one has a temperature $\sim$100~keV
and low optical depth, $\tau\sim 1$, and is responsible for the
high-energy emission (e.g. Zdziarski et al. 1996). The cool region
with higher optical depth smoothly extends the disc emission up to
soft X-ray energies (e.g. Laor et al. 1997). Physically, this
second component could arise e.g. in the transition region between
a disc and optically thin inner flow (Magdziarz et al.~1998) or in
a warm skin on the disc surface (Janiuk, Czerny \& Madejski 2001).
Using the disc as seed photons for a cool Compton scattering cloud
can also explain the strong correlation between optical/UV and
soft X-ray luminosity (e.g. Walter \& Fink 1993; Marshall et
al.~1997) and fits well with the inferred shape of the UV-to-soft
X-ray continuum in quasars (Zheng et al.~1997).

However, changing the disc seed photons should have some impact on
the Comptonization region. An electron cloud of constant optical
depth and heating power, $L_h$, has an equilibrium temperature
which is lower for stronger seed photon illumination, $L_s$ (e.g.
Page et al. 2003). The Comptonization model for the soft excess
makes the clear prediction that the soft excess shape should
correlate with $L_h/L_s$.

We might also expect that the AGN spectra should be analogous to
the Galactic black holes (GBH), modulo whatever changes are
produced by changing the accretion disc temperature. At high
luminosities, the Galactic binary systems have X-ray spectra
consisting of the disc emission and a Comptonized tail of variable
relative intensity. The range of soft-state spectra (from
ultrasoft through high to very high states) seen in GBH requires
changing $L_h/L_s$ by over 2 orders of magnitude (Done \&
Gierli\'{n}ski 2003). Hence we would expect that a sample of soft
state (disc dominated) AGN should show a corresponding range in
shape of the soft X-ray excess.

In this {\it Letter} we quantify the soft excess properties on a
large sample of good quality AGN spectra from the
\emph{XMM-Newton}. We use these to test the Comptonization model
for the origin of the soft X-ray excess.

\section{Data reduction}

We use the bright quasar sample (Boroson \& Green 1992) as a
starting point, as these objects (mainly radio quiet quasars) are
selected by their strong blue/UV continuum flux, i.e. have a
strong accretion disc component. These are all well studied, so
have known (and fairly small) $E(B - V)$ values, together with
good bolometric luminosity and mass estimates (Boroson 2002; Woo
\& Urry 2002). This means their luminosity can be estimated as a
fraction of Eddington luminosity, $L/L_{\rm Edd}$, with the
majority spanning the range between $0.1 < L/L_{\rm Edd} < 1$. The
observed disc dominance of the AGN spectra in this range is as
expected from corresponding $L/L_{\rm Edd}$ spectra from accreting
GBH (e.g. Done \& Gierli{\'n}ski 2003)

From this sample we selected all the publicly available (as in
September 2003) X-ray spectra from \emph{XMM-Newton} archive.
After excluding the radio-loud (PG~0007+106, PG~1226+023 and
PG~1425+267), heavily-absorbed (PG~1114+445 and PG~2214+139) and
very faint (PG~1411+442) objects, we finally obtained a sample of
26 sources, with black hole masses, $M$ and bolometric
luminosities in Eddington units, $L/L_{\rm Edd}$ taken from
Boroson (2002), as shown in Table \ref{tab:log}.

\begin{table}
\begin{tabular}{ccccc}
  \hline
  PG number & $z$ & $\log(M/$M$_{\odot})$ & $L/L_{\rm Edd}$ & Obs. date\\
  \hline
  0003+199 & 0.025  & 7.07 & 0.62 & 2000-12-25 \\
  0050+124 & 0.061  & 7.13 & 1.81 & 2002-06-22 \\
  0157+001 & 0.163  & 8.00 & 0.54 & 2000-07-29 \\
  0804+761 & 0.100  & 8.21 & 0.68 & 2000-11-04 \\
  0844+349 & 0.064  & 7.66 & 0.41 & 2000-11-04 \\
  0947+396 & 0.206  & 8.46 & 0.14 & 2001-11-03 \\
  0953+414 & 0.239  & 8.52 & 0.58 & 2001-11-22 \\
  1048+342 & 0.167  & 8.14 & 0.25 & 2002-05-13 \\
  1115+407 & 0.154  & 7.44 & 0.82 & 2002-05-17 \\
  1116+215 & 0.177  & 8.41 & 0.74 & 2001-12-02 \\
  1202+281 & 0.165  & 8.37 & 0.11 & 2002-05-30 \\
  1211+143 & 0.085  & 7.81 & 1.14 & 2001-06-15 \\
  1244+026 & 0.048  & 6.24 & 3.97 & 2001-06-17 \\
  1307+085 & 0.155  & 8.50 & 0.24 & 2002-06-13 \\
  1309+355 & 0.184  & 8.20 & 0.33 & 2002-06-10 \\
  1322+659 & 0.168  & 7.74 & 0.81 & 2002-05-11 \\
  1352+183 & 0.158  & 8.20 & 0.29 & 2002-07-20 \\
  1402+261 & 0.164  & 7.76 & 1.24 & 2002-01-27 \\
  1404+226 & 0.098  & 6.65 & 1.90 & 2001-06-18 \\
  1426+015 & 0.086  & 8.79 & 0.10 & 2000-07-28 \\
  1427+480 & 0.211  & 7.86 & 0.57 & 2002-05-31 \\
  1440+356 & 0.077  & 7.28 & 1.07 & 2001-12-23 \\
  1444+407 & 0.267  & 8.17 & 0.71 & 2002-08-11 \\
  1501+106 & 0.036  & 8.23 & 0.12 & 2001-01-14 \\
  1613+658 & 0.138  & 9.48 & 0.03 & 2001-08-29 \\
  1626+554 & 0.132  & 8.24 & 0.15 & 2002-05-05 \\
  \hline
\end{tabular}
\caption{List of PG quasars analysed here. Mass and luminosity
were taken from Boroson (2002).} \label{tab:log}
\end{table}

We extracted the X-ray spectra from EPIC MOS1, MOS2 and PN using
{\sc sas} 5.4 software. During the PG~1501+106 observation MOS1
was in timing mode, so only MOS2 data were available. Only X-ray
events corresponding to patterns 0--12 (MOS) and 0--4 (PN) were
used. Source and background spectra were extracted from circular
regions of 45 arcsec in radius around the object and in an offset
position clear of other sources, respectively. For spectral
analysis we used {\sc xspec} software, fitting MOS and PN spectra
simultaneously in the 0.3--10~keV energy band.

\section{Spectral fitting and results}
\label{sec:fitting}

We fit each spectrum with a model consisting of two Comptonization
continua. The hot component produces the power-law spectrum, while
the cool one creates the soft excess. We use {\sc thcomp}
(Zdziarski, Johnson \& Magdziarz 1996) to describe the shape of
these spectra. This is based on an approximate solution of the
Kompaneets (1956) equation, and is parameterized by the asymptotic
photon spectral index and electron temperature. We assume that the
seed photons for both Compton scattering regions come from a disc
blackbody of temperature 10~eV, which is far outside the
0.3--10~keV energy band so has no effect on the observed spectral
shape. We fix the temperature of the hot component at 100~keV, as
its high-energy cutoff is not seen below 10 keV so the temperature
cannot be constrained by the data. The cool component temperature
is well constrained, but its spectral index (or equivalently,
optical depth) is not, so we fix this at $\Gamma_{\rm soft} =
2.0$. This intrinsic spectrum is absorbed by column $N_H$, which
we allow to vary freely.

The soft excess is statistically significant in all the spectra.
The smallest change in $\chi^2$ (for PG~0157+001) is 26 for two
extra degrees of freedom compared to the fits without a cool
component, giving an F-test significance level of $\sim10^{-6}$.
The statistical significance of the soft excess is not necessarily
proportional to its strength as the signal-to-noise ratio varies
from spectrum to spectrum.

\begin{figure*}
\begin{center}
\leavevmode \psfig{file=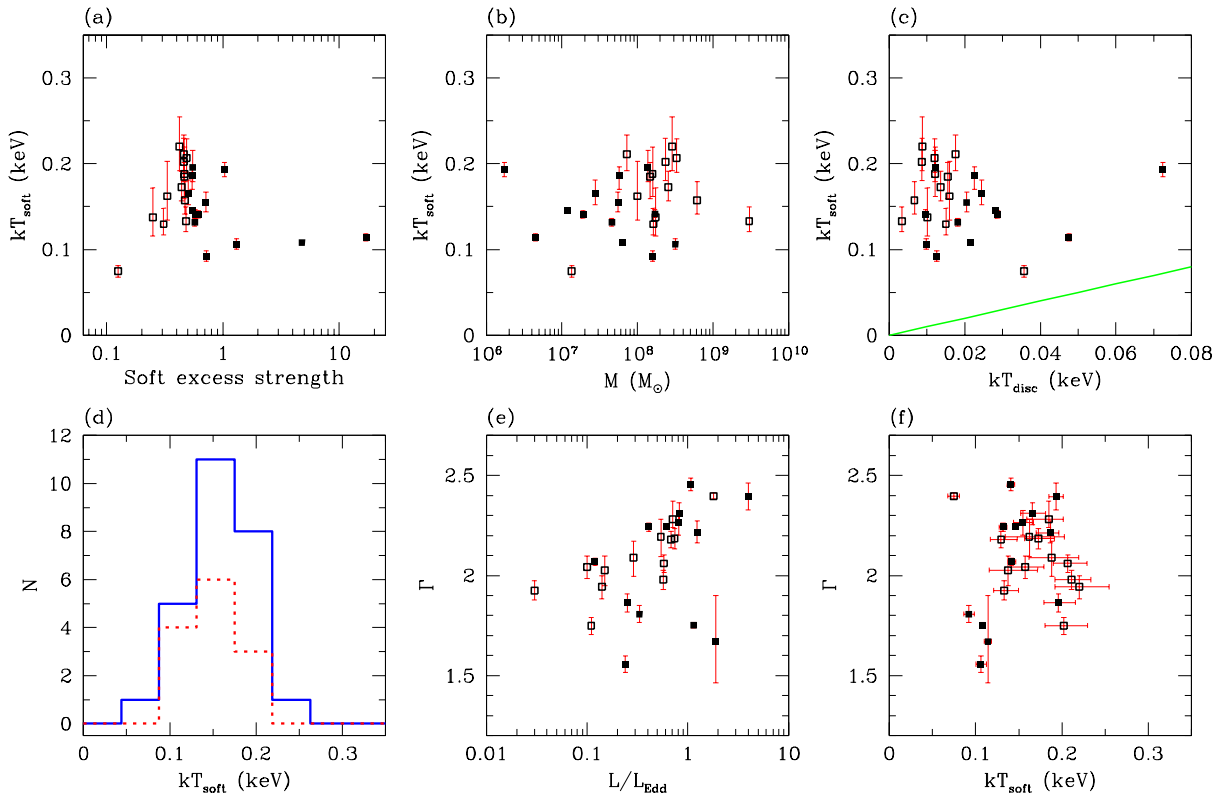,width=0.9\textwidth}
\end{center}
\caption{Spectral fitting results from the model of two
Comptonizing regions. The open and filled symbols correspond to
the soft excess strength, $R_{\rm exc}$, less and greater then
0.5, respectively (see Sec.~\ref{sec:fitting}). (a) Temperature of
the cool component which produces the soft excess, $T_{\rm soft}$,
versus strength of the soft excess. (b) Distribution of $T_{\rm
soft}$ versus black hole mass. (c) Comparison of the estimated
accretion disc temperature, $T_{\rm disc} \propto (L/L_{\rm
Edd})^{1/4} M^{-1/4}$ and observed $T_{\rm soft}$. The line
represents $T_{\rm disc} = T_{\rm soft}$. (d) Histogram of $T_{\rm
soft}$. The solid and dotted lines correspond to the full sample
and sources with $R_{\rm exc} \ge$ 0.5, respectively. (e) Photon
spectral index $\Gamma$ of the hot component versus fraction of
the Eddington luminosity. (f) Spectral index of the hot
Comptonization, $\Gamma$, versus $T_{\rm soft}$.}
\label{fig:results}
\end{figure*}

Fig.~\ref{fig:results}a shows the characteristics of the soft
excess for each object, plotting temperature against strength of
the soft excess, $R_{\rm exc}$, measured by the ratio of
unabsorbed 0.3--2 keV flux in the cool and hot components. The
highest excesses of $R_{\rm exc}$ = 4.7 and 17 is in PG~1211+143
and PG~1404+226, respectively, though the spectrum of the latter
one has poor signal-to-noise.  Other sources have $R_{\rm exc} \la
1$. We use the condition $R_{\rm exc}=0.5$ to crudely divide our
sample into objects with strong and weak excess, denoted by filled
and open symbols respectively in Fig.~\ref{fig:results}.

The most striking property of the soft excess is its constancy in
temperature. It is distributed in a very narrow range of values
between 0.1 and 0.2~keV (Fig.~\ref{fig:results}a), with tendency
to concentrate close to 0.1~keV for sources with large $R_{\rm
exc}$. The mean temperature is $\langle kT_{\rm soft}\rangle =
0.12$~keV and variance $\sigma = 0.02$~keV. Even more interesting
is the independence of $T_{\rm soft}$ on the black hole mass
(Fig.~\ref{fig:results}b). The maximum temperature of the standard
Shakura-Sunyaev disc is $T_{\rm disc} \propto M^{-1/4} (L_{\rm
disc}/L_{\rm Edd})^{1/4}$, so we can estimate it from the values
given in Table 1. Applying a spectral hardening factor of 1.8
(Shimura \& Takahara 1995) we find that the expected disc
temperature should be between $\sim$3 and $\sim$70 eV. This is a
rough estimate only, as masses and luminosities are uncertain, as
is the fraction of the total luminosity released in the disc.
Nevertheless, two things are clear (see Fig.~\ref{fig:results}c).
Firstly, the observed soft excess temperatures of $\sim$0.1~keV
are too high to be direct disc emission. Secondly, the range of
variation in $T_{\rm soft}$ (about a factor 2) is much smaller
than the range in disc temperatures (about factor 20). Thirdly,
there is no correlation between $T_{\rm disc}$ and $T_{\rm soft}$.

The hot component, represented here by Comptonization of the disc
photons, is simply a power law in the 0.3--10~keV band, since the
seed photons are well below and the high-energy rollover is well
above the observed energy band. Its photon spectral index,
$\Gamma$, varies between $\sim$1.5 and $\sim$2.5. There is a weak
positive correlation between $\Gamma$ and luminosity $L/L_{\rm
Edd}$ (see Fig.~\ref{fig:results}e), though unlike the X-ray
binaries, there are bright sources ($L/L_{\rm Edd} > 0.1$) with
hard spectra ($\Gamma < 2$). In particular, PG~1211+143 has
$\Gamma \approx 1.75$ near Eddington luminosity. There is no
obvious correlation between the soft excess temperature and
$\Gamma$ (Fig.~\ref{fig:results}f).

The model generally gives a good fit to the spectra, with reduced
$\chi^2/\nu < 1.3$ for most of the data. However, in a few of the
highest signal-to-noise spectra there are clear narrow residual
features from absorption which are not modelled by our continuum
fits. These are most noticeable for PG~1211+143, where $\chi^2/\nu
= 2.0$. Detailed studies of its {\em XMM-Newton} spectrum has
revealed the presence of complex ionized absorption (Pounds et
al.~2003). Adding multiple absorbers, together with ionized
reflection from the accretion disc, improves the fit ($\chi^2/\nu
= 1.36$) but only weakly affects the underlying continuum. In
particular, the strength and temperature of the soft excess remain
very similar.

\section{The Nature of the soft excess}

Comptonization of the accretion disc spectrum can successfully fit
the shape of the soft X-ray excess below $\sim$2~keV in all these
Seyfert 1/radio-quiet quasars. However, the temperature of this
putative Comptonizing region remains remarkably constant at
0.1--0.2 keV over all 26 different objects, which have a large
range in mass and luminosity, hence in the disc temperature and,
most likely, ratio of the power released in hot plasma to that of
the disc.

We use the {\sc eqpair} Comptonization code (Coppi 1999) to
quantify possible changes in the Comptonized spectrum. The key
advantage of this code is that it does not assume the steady-state
electron distribution, rather it {\em calculates} it by balancing
heating (primarily the injected power in the hot electrons, $L_h$)
and cooling (proportional to the power in seed photons, $L_s$).
The spectral shape is mostly determined by the ratio $L_h/L_s$,
which is dependent on the intrinsic mechanisms and geometry of the
source.

We expect the disc temperature to vary by factor $\sim$20 within
our sample (see Fig.~\ref{fig:results}c). This translates into
variation by factor $\la$3 in the equilibrium electron temperature
of the Comptonizing plasma with constant $L_h/L_s$ and optical
depth. Thus, a large variation in the seed photon temperature can
lead to only a moderate change in the electron temperature, and it
seems feasible that the soft excess can be produced by a cool,
optically thick, Comptonized component.

This however, requires constancy of $L_h/L_s$ and -- to lesser
extent -- optical depth. Large variations in $L_h/L_s$ would lead
to significant change in both index and temperature of the
Comptonized spectrum, which is inconsistent with our measurements.
Constancy of $L_h/L_s$ could be explained e.g. if the seed photons
were dominated by reprocessed flux in a constant (disc-corona?)
geometry. Nonetheless the hot Comptonized component {\em is}
variable. The observed span in its spectral index (see
Fig.~\ref{fig:results}e), between $\sim$1.5 and $\sim$2.5 {\em
requires} a corresponding change in $L_h/L_s$ from $\sim$10 to
$\sim$0.5. Thus the two Comptonizing regions have to be spatially
separate and independent.

This constancy of soft spectral shape is even more puzzling when
compared to GBH, which show a wide variety of spectral shapes for
$0.1<L/L_{\rm Edd}<1$ (e.g. Done \& Gierli\'{n}ski 2003). If the
accretion properties simply scale up with the black hole mass, we
should see a similar variety in AGN, yet we don't. The soft-state
spectra of GBH can be modelled by Comptonization on two electron
distributions: the cooler thermal and the hotter power law (e.g.
Gierli{\'n}ski \& Done 2003, Zdziarski et al. 2001). The thermal
electrons create a soft `hump' above the power law, but their
derived moderate optical depth is much smaller than $\tau \sim$
50, required here. This gives rise to a much smoother and broader
shape than the observed soft excess in e.g. PG~1211+143.

\section{Alternative solutions}

The universality of the soft excess shape has been noticed before
(e.g. Walter \& Fink 1993; Czerny et al. 2003). It seems to be
much simpler to explain if it is set by some characteristic,
physical energy of the system, such as given by atomic
transitions. One obvious feature in the soft energy band is the
the strong jump in opacity associated with lines and edges of
ionized {\sc o\,vii}, {\sc o\,viii} and iron at $\sim$0.7 keV.
This could have an impact either through absorption or through
reflection.

The reflection probability depends on the balance between electron
scattering and photoelectric absorption opacity, so the decrease
in opacity below $\sim$0.7~keV causes a strong increase in the
reflected fraction below this energy for an ionized disc. Czerny
\& {\.Z}ycki (1994) showed that such ionized reflection from a
disc could fit the moderate resolution {\it ROSAT} data on the
soft excess in some AGN, though they required a fairly high
reflected fraction, with $\Omega/2\pi \sim 2$. Higher resolution
data from {\it XMM-Newton} show no obvious spectral features at
$\sim$0.7~keV, so if atomic features are responsible for the soft
excess then they must be strongly smeared by high velocities
and/or gravitational redshifts (Ross, Fabian \& Ballantyne 2002).
Inner disc, ionized reflected spectra can fit the strong soft
excesses seen in the {\em XMM-Newton} spectrum of 1H~0707--495 and
MCG-6-30-15 (Fabian et al. 2002; Ballantyne et al. 2003), although
the models require a puzzling range of ionization states, and a
geometry in which the direct emission from the hard X-ray source
is hidden.

\begin{figure}
\begin{center}
\leavevmode \psfig{file=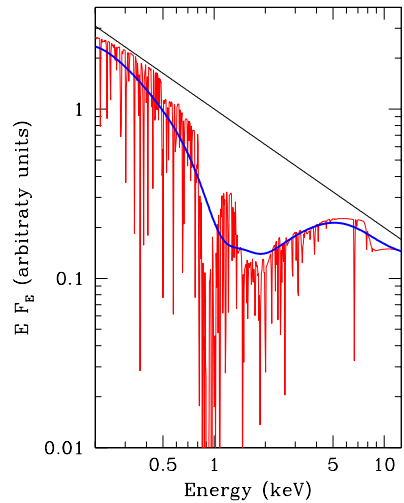,width=0.3\textwidth}
\end{center}
\caption{An illustration of the ionized absorption model creating
the soft excess. The upper line is an underlying power law with
spectral index $\Gamma$ = 2.7. The red line (grey in B\&W) shows
the multiple absorption features predicted by an {\sc xstar} model
with $N_H = 33\times10^{22}$~cm$^{-2}$ and $\xi$ = 460 erg cm
s$^{-1}$ (best-fitting parameters to PG~1211+143). The curved
thick line shows the this absorbed continuum convolved with a
Gaussian of $\sigma/E$ = 0.28. This figure can be seen in colour
in the on-line version of this article on {\it Synergy}.}
\label{fig:smodel}
\end{figure}

Here we propose instead that the soft excess arises from
relativistically smeared {\em absorption}, where the
characteristic atomic spectral features are again masked by high
velocities. Physically, this absorption could arise from a
differentially rotating, outflowing disc wind (Murray \& Chiang
1997). The complex velocity structure of such a wind is beyond the
scope of this {\it Letter}, but we roughly model it by a Gaussian
velocity dispersion with width $v/c\sim 0.2$ convolved with {\sc
xstar} (Bautista \& Kallman 2001) model of the ionized absorption
column. This smeared absorption produces a smooth `hole' in the
spectrum which results in an apparent soft excess at low energies,
and hardens the spectrum at higher energies (see
Fig.~\ref{fig:smodel}). The smearing is so large that all the
features, including the iron line edge at $\sim$7~keV, are removed
from the absorption spectrum, leaving a smooth continuum (solid
thick curve in Fig.~\ref{fig:smodel}).

Using this complex absorption, we can fit the spectrum of
PG~1211+143 with a single Comptonized continuum, without requiring
an additional cool Comptonized component to model the soft excess.
Including the additional {\em narrow} absorption components and
reflection from the disc ($\Omega/2\pi \sim 1$) gives a slightly
worse fit than with the two-component model, with $\chi^2/\nu =
1.65$. However, the residuals are concentrated at $\sim 0.6$~keV,
consistent with {\em emission} from {\sc o\,vii}/{\sc o\,viii} in
the wind, or perhaps indicating the inadequacies of a Gaussian
velocity profile. Apart from these details, the overall continuum
shape is well matched by this model. We leave detailed analysis of
this and other sources for future work.

\begin{figure}
\begin{center}
\leavevmode \psfig{file=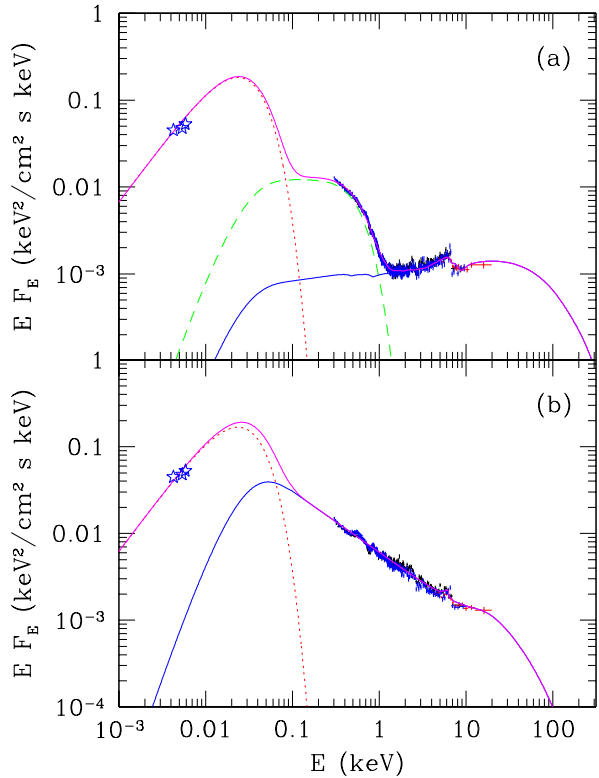,width=0.45\textwidth}
\end{center}
\caption{Unabsorbed model components together with unfolded and
unabsorbed data from PG~1211+143. Open stars represent OM data,
black, blue and red crosses represent EPIC MOS, PN and
(non-simultaneous) {\it RXTE} PCA data, respectively. The dotted
red curve represents the disc spectrum. (a) The model with two
Comptonized components and ionized reflection. In this model the
soft excess comes from Comptonization in warm optically thick
plasma (dashed green curve). (b) The model with complex ionized,
relativistically smeared absorption, which is not seen here as
this is \emph{unabsorbed} model. Instead, the intrinsic spectrum
is very different and does not require any additional component to
explain the soft excess. In this interpretation the soft excess is
solely due to complex absorption. This figure can be seen in
colour in the on-line version of this article on {\it Synergy}.}
\label{fig:unspec}
\end{figure}

Fig.~\ref{fig:unspec} shows a comparison of the deconvolved and
unabsorbed spectra of PG~1211+143 for the two models of the soft
excess. The model with a separate Comptonization component
(Fig.~\ref{fig:unspec}a) has a strong soft excess, with a hard
spectrum at high energies ($\Gamma\sim 1.8$).  By contrast, for
the smeared, ionized absorber, there is only a single continuum
component, with much steeper spectral index ($\Gamma\sim 2.7$).
This is much closer to what is seen in the high luminosity GBH
systems, which generally have $\Gamma > 2$.

%=================================================================

\section{Conclusions}

The soft X-ray excess in radio quiet quasars can be modelled by
including a second, cool Comptonizing component. However, the
observed range in its temperature is very small, despite large
changes in black hole mass, luminosity and 2-10~keV X-ray spectral
index across the sample. This requires some fine tuning of the
parameters, and strongly contrasts with the analogous soft-state
GBH, with large diversity of the X-ray spectral shape over a
comparable range in $L/L_{\rm Edd}$.

Instead, we propose that the soft excess is an artifact of ionized
absorption in a wind from the inner disc. The strong jump in
opacity at $\sim$0.7~keV seen from {\sc o\,vii}, {\sc o\,viii} and
iron in moderately ionized material can give an apparent soft
excess whose energy is fixed by atomic physics, so trivially is
constant across a wide range of objects. If this absorber is
associated with an accretion disc wind then its complex velocity
structure (differentially rotating and outflowing) gives rise to
substantial {\em broadening} which masks the sharp atomic
features, leaving a smooth continuum. In PG~1211+143 we show that
such models can fit the data {\em without} requiring a separate
soft excess. This can dramatically change the inferred intrinsic
spectral slope, which has obvious implications for the robustness
of the derived parameters for the broad iron line from the disc.
The much softer intrinsic spectrum is now consistent with the soft
continua seen from high-luminosity GBH. It also means that these
two models for the origin of the soft excess can be unambiguously
tested with high signal-to-noise, broad bandpass (0.1--50~keV)
data, such as will be obtained by {\it ASTRO-E2}.

\section*{Acknowledgements}

We thank Todd Boroson for the black hole data.

\label{lastpage}

\end{document}